\documentclass[aps,prl,twocolumn,showpacs]{revtex4}
\usepackage{amssymb}

\usepackage{graphicx}

\begin{document}

\title{Hybridization of spin and plasma waves in Josephson tunnel junctions
with a ferromagnetic layer. }
\author{A. F. Volkov${}^{1,2}$}
\author{K. B. Efetov${}^{1}$.}

\address{$^{(1)}$Theoretische Physik III,\\
Ruhr-Universit\"{a}t Bochum, D-44780 Bochum, Germany\\
$^{(2)}$Institute for Radioengineering and Electronics of Russian Academy of\\
Sciences,11-7 Mokhovaya street, Moscow 125009, Russia\\}

\begin{abstract}
We study dynamics of tunnel Josephson junctions with a thin ferromagnetic
layer F [superconductor-insulator-ferromagnet-superconductor (SIFS) junctions].
 On the basis of derived equations relating the
superconducting phase and magnetic moment to each other we analyze
collective excitations in the system and find a new mode which is a hybrid
of plasma-like and spin waves. The latter are coupled together in a broad
range of parameters characterizing the system. Using the solution describing
the collective modes we demonstrate that besides the Fiske steps new peaks
appear on the I-V characteristics due to oscillations of the magnetic moment
$M$ in the ferromagnetic layer. Thus, by measuring the I-V curve of the SIFS
junctions, one can extract an information about the spectrum of spin
excitations in the ferromagnet F.
\end{abstract}

\maketitle

\bigskip

Discovery of the Josephson effects in 1962 was an important step in the
development of the condensed matter physics \cite{Josephson}. Josephson has
predicted that the dc current of Cooper pairs with a magnitude of order of
the qusiparticle current can flow in the absence of the voltage\ through a
thin insulating layer separating two superconductors S. In the presence of a
voltage $V$\ the supercurrent oscillates in time with the Josephson
frequency $\omega _{J}=2eV/\hbar $.

On the basis of equations describing electrodynamics of the Josephson tunnel
junction [superconductor-insulator-superconductor (SIS)] many fascinating phenomena were predicted \cite%
{Josephson,Barone}. For example, small perturbations of the phase difference
$\varphi $\ may propagate in SIS junctions in a form of Josephson
\textquotedblleft plasma\textquotedblright\ waves with the spectrum $\omega
^{2}=\Omega _{J}^{2}(1+(kl_{J})^{2}).$\ Large perturbations exist in a form
of solitons carrying the magnetic flux quantum (fluxons or antifluxons). The
interaction of the \textquotedblleft plasma\textquotedblright\ waves with
oscillating Josephson currents leads to resonances and peculiarities on the
I-V characteristics (Fiske steps) \cite{Barone}.

The Josephson junctions (JJ) based on conventional superconductors are
widely used in practice as generators, the most sensitive detectors of
magnetic fields and ac radiation in a wide range of the frequency specrtum,
etc \cite{Barone}. Recently, a considerable progress has been achieved in
practical applications of the JJs based on high $T_{c}$\ superconductors %
\cite{Koshelev}.

Replacing the insulating (I) layer by a ferromagnet (F) one obtains SFS
junctions that have been under intensive study during the last decade \cite%
{GolubovRMP,BuzdinRMP,BVErmp,Lyuksyutov}. New interesting effects like the
so-called $\pi -$state (negative critical\ Josephson current $j_{c}$) or new
type of superconducting correlations (odd triplet superconductivity) may
arise in JJs of this type \cite{BuzdinRMP,BVErmp,EschrigR}. Although the
main attention was paid to the study of the dc Josephson effect, some ac
effects in SFS or SmS (m means a magnetic nanoparticle) junctions have been
considered, too \cite{acJos,Braude,Maekawa}.

At the same time, all these studies\textbf{\ }deal with the SFS junctions
where effects like those in tunnel JJs are absent. Experimental studies of
the SIFS junctions with an insulating layer I have started quite recently
(see \cite{Weides} and references therein). In these tunnel JJs one can
observe collective modes and other dynamical phenomena. Note that the effect
of superconductivity on the ferromagnetic resonance has been studied
experimentally on S/F bilayers \cite{ExpDyn}. We have no doubts that proper
experiments will be performed in the nearest future and theoretical
description of dynamical effects in SIFS or SFIFS tunnel junctions is in
great demand.

By now, no theory for dynamical effects in SIFS or SFIFS tunnel junctions
has been suggested, although one can expect really new effects. One of the
basic properties of the JJs is that they are very sensitive to even very
small variations of the magnetic field. So, one can easily imagine that
dynamics of the spin degrees of freedom leading to variations of the
magnetic field should essentially affect electrodynamics of tunnel SIFS
junctions.\textbf{\ }

In this paper, we develope theory describing dynamical effects in SIFS or
SFIFS junctions. We demonstrate that dynamics of the phase difference $%
\varphi $ and of the magnetization $M$ is described by two coupled equations
which determine dynamics of both the spin and orbital degrees of freedom.
One of these equations governs the spatial and temporal evolution of the
phase $\varphi $, while the other describes the precession of the
magnetization $M.$ Using these equations we investigate collective modes in
Josephson junctions with F layers and show that these modes are coupled spin
and plasma-like Josephson waves.

Although the influence of the superconducting condensate on spin waves in
magnetic superconductors was studied by Braude and Sonin \cite{Sonin} some
time ago, the tunnel JJs with a ferromagnetic layer is a more interesting
system because besides the spin waves another mode (plasma-like Josephson
waves) exists in these systems. It is our main finding that in such systems
the both types of the excitation hybridize forming a new collective mode
consisting of coupled oscillations of the magnetization and the Josephson
current\textbf{.} Moreover, the spin waves can be excited and recorded
simply by passing a dc current $j$ through the junctions, and measuring the
I-V characteristics of the junction. We demonstrate below that peculiarities
on the I-V curve (Fiske steps) bear information about the spectrum of the
system, which can be considered as the basis of a new type of spectroscopy
of spin excitations in a ferromagnet\textbf{. }

The hybridization of the charge and spin excitations occurs because the
contribution to the magnetic moment in the system comes from both orbital
electron motion and exchange field. The orbital currents are affected by
spins of the ferromagnets and vice versa. The coupling between the
plasma-like and spin waves in SIFS junctions may show up in additional
resonances and peaks on the I-V characteristics.

Having discussed the physics of the SIFS junctions qualitatively, let us
present now the quantitative theory. We consider for simplicity an SFIFS
junction (the obtained results are valid also for an SIFS junction shown
schematically in Fig.1a) and calculate the in-plane current density $%
j_{\perp }$ as a function of the magnetic induction $B_{\perp }$ in the
plane parallel to the interface. This current can be expressed through the
vector potential $A_{\perp }$ \textbf{(}$\mathbf{\nabla \times A=B}$\textbf{)%
} and gradient of the phase $\nabla _{\perp }\chi $ as
\begin{equation}
\mathbf{j}_{\perp }=(c/4\pi )\lambda _{L}^{-2}(-\mathbf{A}_{\perp }+(\Phi
_{0}/2\pi )\mathbf{\nabla }_{\perp }\chi )  \label{a1}
\end{equation}%
where $\lambda _{L}$ is the London penetration length and $\Phi _{0}=hc/2e$
is the magnetic flux quantum.

Writing Eq. (\ref{a1}) we imply a local relationship between the current
density $\mathbf{j}_{\perp }$ and the gauge invariant quantity in the
brackets, which is legitimate in the limit $k\lambda _{L}\ll 1$, where $k$
is the modulus of the in-plane wave vector of perturbations. Subtracting the
expressions for the current density, Eq. (\ref{a1}), written for the right
(left) superconductors from each other we find a change of the current
density $[\mathbf{j}_{\perp }]$ across the junction

\begin{figure}[tbp]
\begin{center}
\includegraphics[width=4cm, height=5cm]{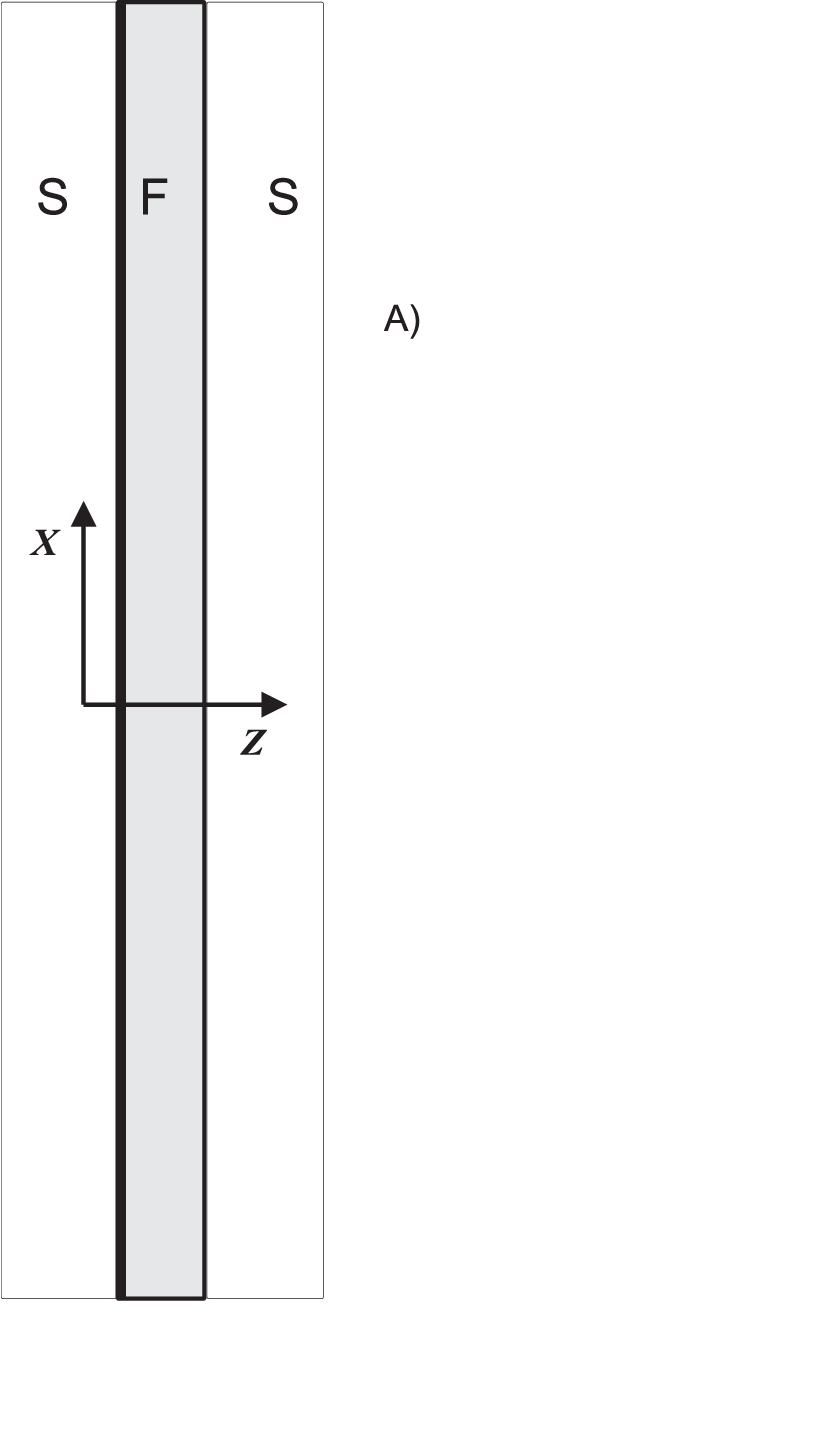} %
\includegraphics[width=4.5cm, height=5cm]{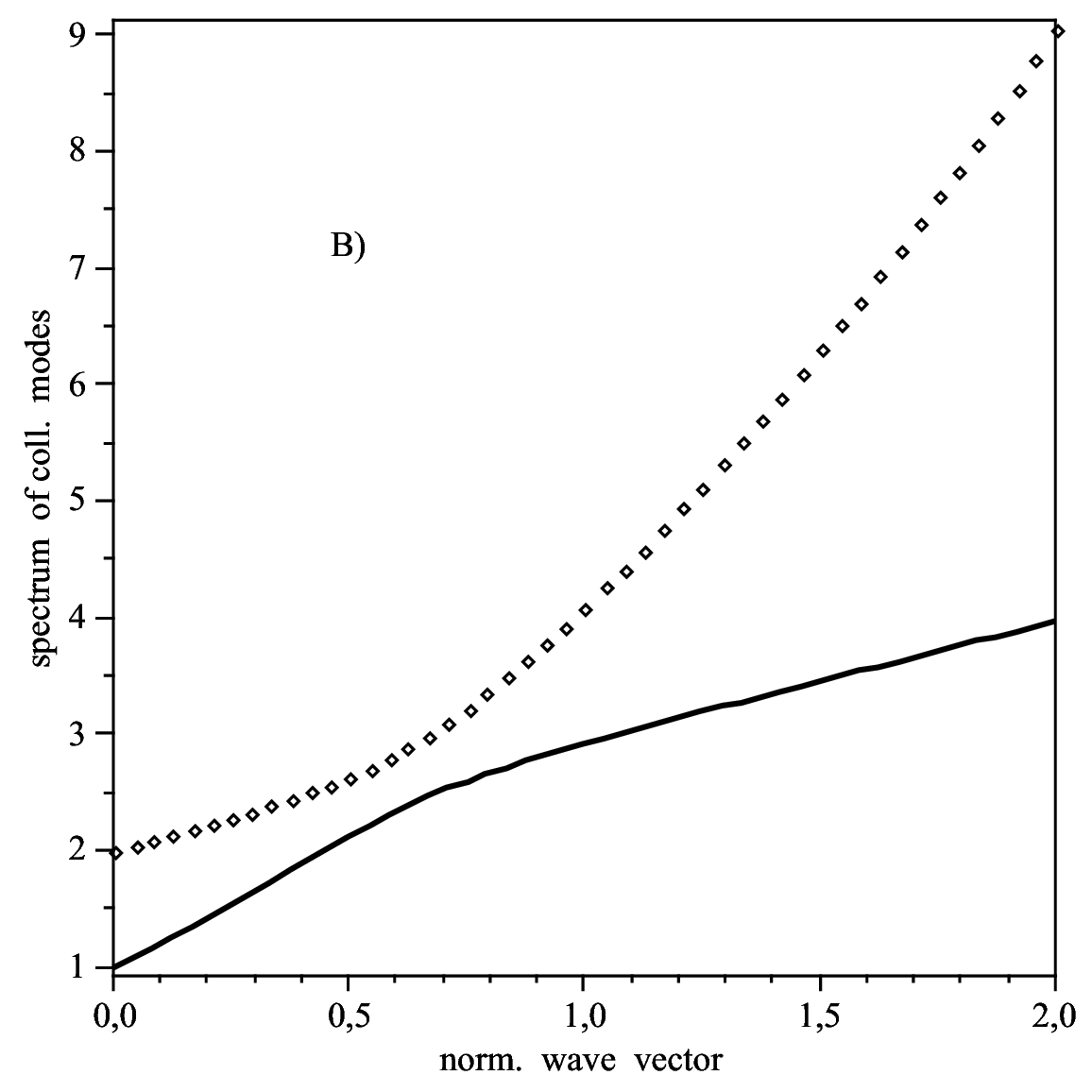}
\end{center}
\caption{A) Schematic picture of the system under consideration. The thick
line between the S and F layers means an insulating layer; B) Spectrum of
coupled spin and plasma-like modes. On the vertical and horizontal axes we
plot the quantity $(\protect\omega/\Omega_{J})^2$, and $(kl_{J})^2$,
respectively. The following parameters are chosen: $(\Omega_{J}/%
\Omega_{M})^2=2$, $(l_{J}/l_{M})=1$ and $s=0.05$ }
\end{figure}

\begin{equation}
\lbrack \mathbf{j}_{\perp }]=(c/4\pi )\lambda _{L}^{-2}\{4\pi \tilde{d}_{F}(%
\mathbf{n}_{z}\mathbf{\times M}_{\perp })+(\Phi _{0}/2\pi )\mathbf{\nabla }%
_{\perp }\varphi \}  \label{currentJump}
\end{equation}%
where $\tilde{d}_{F}=d_{F}$ (or $2d_{F}$) in the case of a SIFS (or SFIFS)
junction, $d_{F}$ is the thickness of the F film which is assumed to be
smaller than the London penetration length $\lambda _{L}$. This assumption
allows one to neglect the change of $\mathbf{A}_{\perp }$ caused by Meissner
currents in the F layer\textbf{\ }and to write the change of the vector
potential $\mathbf{A}_{\perp }$ in the form $[\mathbf{A}_{\perp }]=\tilde{d}%
_{F}(\mathbf{n}_{z}\mathbf{\times \mathbf{B}_{\perp }})$ with $\mathbf{%
\mathbf{B}_{\perp }}=4\pi \mathbf{M}_{\perp }+\mathbf{H}_{\perp }$.

We proceed by solving the London equation in the superconductors with
boundary conditions determined by the change of the current density (see,
e.g., \cite{AV}). A solution of this equation for S films with the thickness
exceeding $\lambda _{L}$ takes the form

\begin{equation}
\mathbf{B}_{\perp }(z)=\{\frac{\Phi _{0}}{4\pi \lambda _{L}}\mathbf{n}_{z}%
\mathbf{\times \nabla }_{\perp }\varphi -\frac{2\pi \tilde{d}_{F}}{\lambda
_{L}}\mathbf{M}_{\perp }\}\exp (-\frac{(|z|-d_{F})}{\lambda _{L}}).
\label{Induction}
\end{equation}%
Eq. (\ref{Induction}) describes the penetration of the magnetic induction
into the superconductors ($|z|>d_{F}$).\textbf{\ }It is well known that the
magnetic field penetrates the JJ in a form of fluxons \cite{Josephson,Barone}%
. Eq. (\ref{Induction}) supports this picture for the SFIFS junctions.
Integrating Eq. (\ref{Induction}) over $z$ and $x$ and adding the magnetic
flux in the F film $\Phi _{F}=4\pi \tilde{d}_{F}L_{x}M_{y}$ (the magnetic
moment $M_{\perp }$ is assumed to be oriented in the $y$ direction), we come
to flux quantization: $\Phi =$ $(\Phi _{0}n),$ where $n=[\varphi ]_{x}/2\pi $
is the number of fluxons and $[\varphi ]_{x}$ is the phase variation on the
length $L_{x}$, which is supposed to be an integer multiple of $2\pi .$

In order to obtain an equation for the phase difference $\varphi $, we use
the Maxwell equation $(\mathbf{\nabla \times B)}_{z}=(4\pi /c)j_{z}$ in the
superconductors and the standard expression for the Josephson current. Then%
\textbf{, }we obtain

\begin{equation}
\Omega _{J}^{-2}(\frac{\partial ^{2}\varphi }{\partial t^{2}}\mathbf{+}%
\gamma _{J}\frac{\partial \varphi }{\partial t})-l_{J}^{2}\mathbf{\nabla }%
_{\perp }^{2}\varphi +\sin \varphi =\eta -\frac{c\tilde{d}_{F}}{2\lambda
_{L}j_{c}}(\mathbf{\nabla \times M}_{\perp })_{z}  \label{Phase}
\end{equation}%
where $\Omega _{J}=(2ej_{c}/C_{\square }\hbar )^{1/2}$ is the Josephson
``plasma''\ frequency, $\gamma =(R_{\square }C_{\square })^{-1}$, $%
C_{\square }=\epsilon /4\pi d$ and $R_{\square }$ are the capacitance and
resistance of the junction per unit area, $d$ is the thickness of the
insulating layer, $l_{J}^{2}=v_{J}^{2}/\Omega _{J}^{2},$ $v_{J}=c\sqrt{%
d/2\epsilon \lambda _{L}}$ is the velocity of the plasma wave propagation
(Swihart waves). The first term on the r.h.s. of Eq. (\ref{Phase}) , $\eta
=j/j_{c}$, is the normalized current through the junction. A simpler
equation for the stationary case has been written previously in Ref. \cite%
{AV}, where a similar system with a multidomain ferromagnet was considered.

The dynamics of the magnetization in the F layer is described by the well
known equation with account for the magnetic induction due to the Meissner
currents (see for example Ref. \cite{Sonin})

\begin{equation}
\frac{\partial \mathbf{M}_{\perp }}{\partial t}=-4\pi g\alpha (1-\tilde{l}_{M}^{2}%
\mathbf{\nabla }_{\perp }^{2})(\mathbf{M\times M}_{\perp })+g\mathbf{M\times
B}_{\perp }  \label{Mmoment}
\end{equation}%
where $g$ is the gyromagnetic factor, $\alpha $ is a parameter related to
the anisotropy constant $\beta =(\alpha -1)$, $\tilde{l}_{M}$ is a
characteristic length related to the spin waves.

We further substitute $\mathbf{B}_{\perp F}=4\pi \mathbf{M}_{\perp }+\mathbf{%
B}_{\perp S}(d_{F})$ into Eq. (\ref{Mmoment}), where $\mathbf{B}_{\perp F}$
is the magnetic induction in the F layer and $\mathbf{B}_{\perp S}=\mathbf{B}%
_{\perp }(z\rightarrow d_{F})$ is given by Eq.(\ref{Induction}). Finally we
come to the equation

\begin{equation}
-\frac{\partial \mathbf{M}_{\perp }}{\partial t}=\Omega _{M}[\frac{\mathbf{%
M\times M}_{\perp }}{M_{0}}(1+s-l_{M}^{2}\mathbf{\nabla }_{\perp }^{2})+%
\frac{\Phi _{0}}{(4\pi )^{2}\beta \lambda _{L}}\mathbf{\nabla }_{\perp
}\varphi ]  \label{Precession}
\end{equation}%
where $\Omega _{M}=4\pi gM_{0}\beta =4\pi gM_{0}(\alpha -1)$ is the
resonance frequency of the magnetic moment precession ($\alpha >1$), $s=%
\tilde{d}_{F}/(2\beta \lambda _{L}),$ $l_{M}^{2}=(\alpha
/(\alpha -1))\tilde{l}_{M}^{2}$. One could also take into account a damping adding to the
r.h.s. of Eq. (\ref{Precession}) the term $\gamma _{M}(\mathbf{M}\times
\partial \mathbf{M}/\partial t)/M$ ($\gamma _{M}$ is the dimensionless
Gilbert constant).

Eqs. (\ref{Phase}, \ref{Precession}) are the final equations fully
describing dynamics of the SFIFS junctions. The most interesting effect that
we will obtain now is the hybridization of charge and spin excitations
resulting in a new collective hybridized mode.

\textit{a) Hybridized mode. }Let us consider the simplest monodomain case
when the magnetization $\mathbf{M_{0}}$ is normal to the interface, so that
in equilibrium $B_{0}=0$. Small perturbations near the equilibrium result in
a\textbf{\ }precession of the magnetic moment\textbf{\ }$\mathbf{M}$ and in
a variation of\textbf{\ }the phase difference $\varphi $ in time and space.
Representing $\mathbf{M}$ for small deviations\textbf{\ }$\mathbf{m}_{\perp
} $ as $\mathbf{M=}M_{0}\mathbf{n}_{z}+\mathbf{m}_{\perp }$, we can
linearize Eqs. (\ref{Precession}) and (\ref{Phase}). A nonzero solution of
these linearized equations exists provided the determinant of these two
equations equals zero. Writing the perturbations in the form of plane waves (%
$\mathbf{m}_{\perp }\sim \varphi \sim \exp (i\omega t+i\mathbf{kr}_{\perp
})) $ and setting the determinant to zero we come to the dispersion relation

\begin{equation}
\lbrack \omega ^{2}-\Omega _{M}^{2}(k)][\omega ^{2}-\Omega
_{J}^{2}(k)]=s\Omega _{M}(k)\Omega _{M}v_{J}^{2}k^{2}  \label{Drelation}
\end{equation}%
where $\Omega _{M}^{2}(k)=\Omega _{M}^{2}(1+s+(kl_{M})^{2})^{2},$ $\Omega
_{J}^{2}(k)=\Omega _{J}^{2}(1+(kl_{J})^{2}).$\ For simplicity, we neglected
the damping setting $\gamma _{M}\rightarrow 0.$

Eq. (\ref{Drelation})\ describes the spectrum of the hybridized mode. This
mode decouples into the spin and charge excitations only in the limit when
the right hand side can be neglected. In this case the spin waves with the
spectrum $\Omega _{M}(k)$ and the plasma-like Josephson waves with the
spectrum $\Omega _{J}(k)$ exist separately$.$ In the general case we have
the coupled spin wave and plasma-like modes in the system. The most
interesting behavior corresponds to the case $\Omega _{M}<\Omega _{J}$. In
this situation, the two branches of the spectrum cross each other in the
absence of the coupling and the coupling leads to a mutual repulsion of
these branches. In Fig. 1b we show the dependence $\omega (k)$ just for this
case. Note that the frequency of the considered collective modes remains
unchanged by inversion of the magnetization direction, $M\leftrightarrow -M$%
, that is, in a multidomain case we would obtain the same dependence\textbf{%
\ }$\omega (k).$\textbf{\ }

\begin{figure}[tbp]
\begin{center}
\includegraphics[width=4cm, height=5cm]{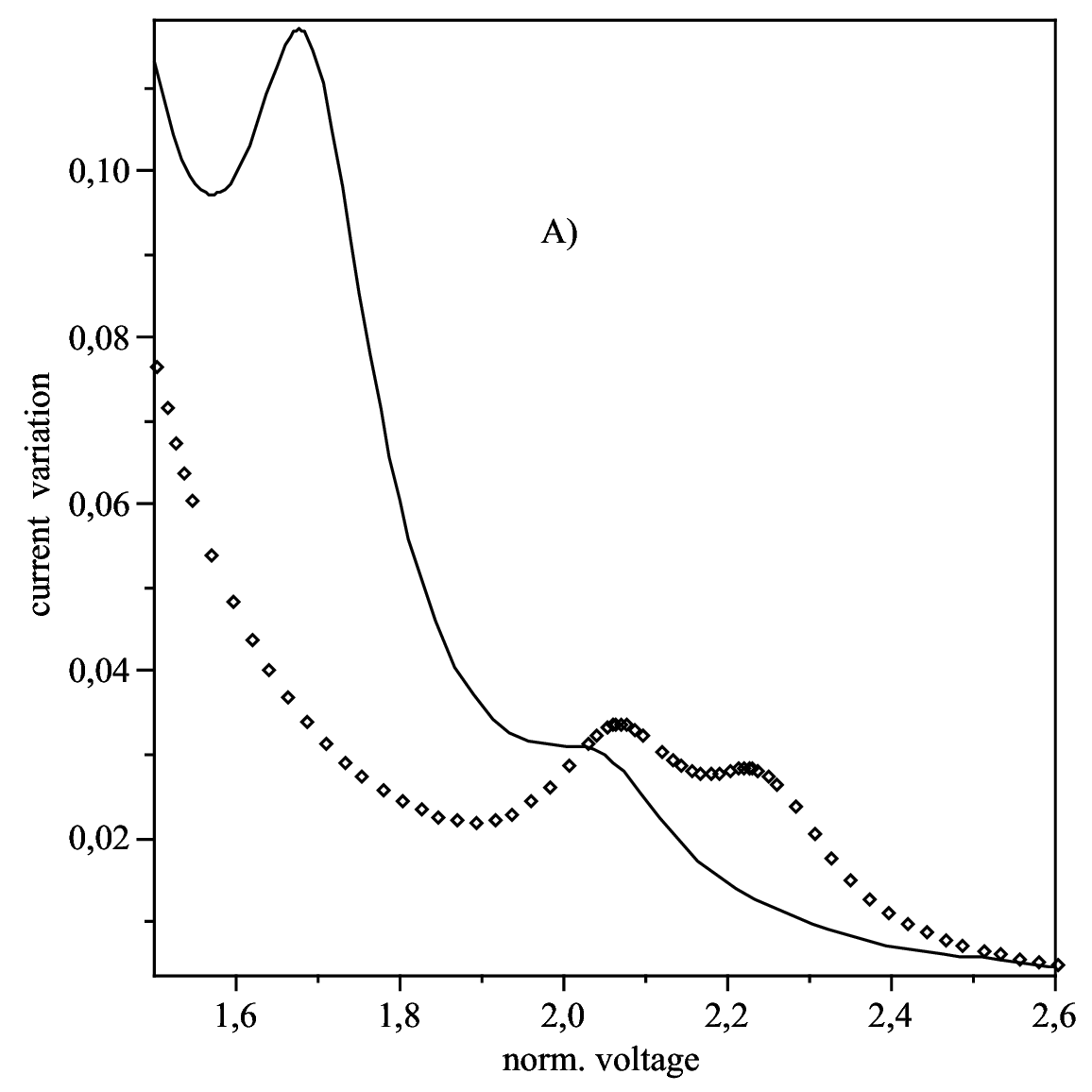} %
\includegraphics[width=4cm, height=5cm]{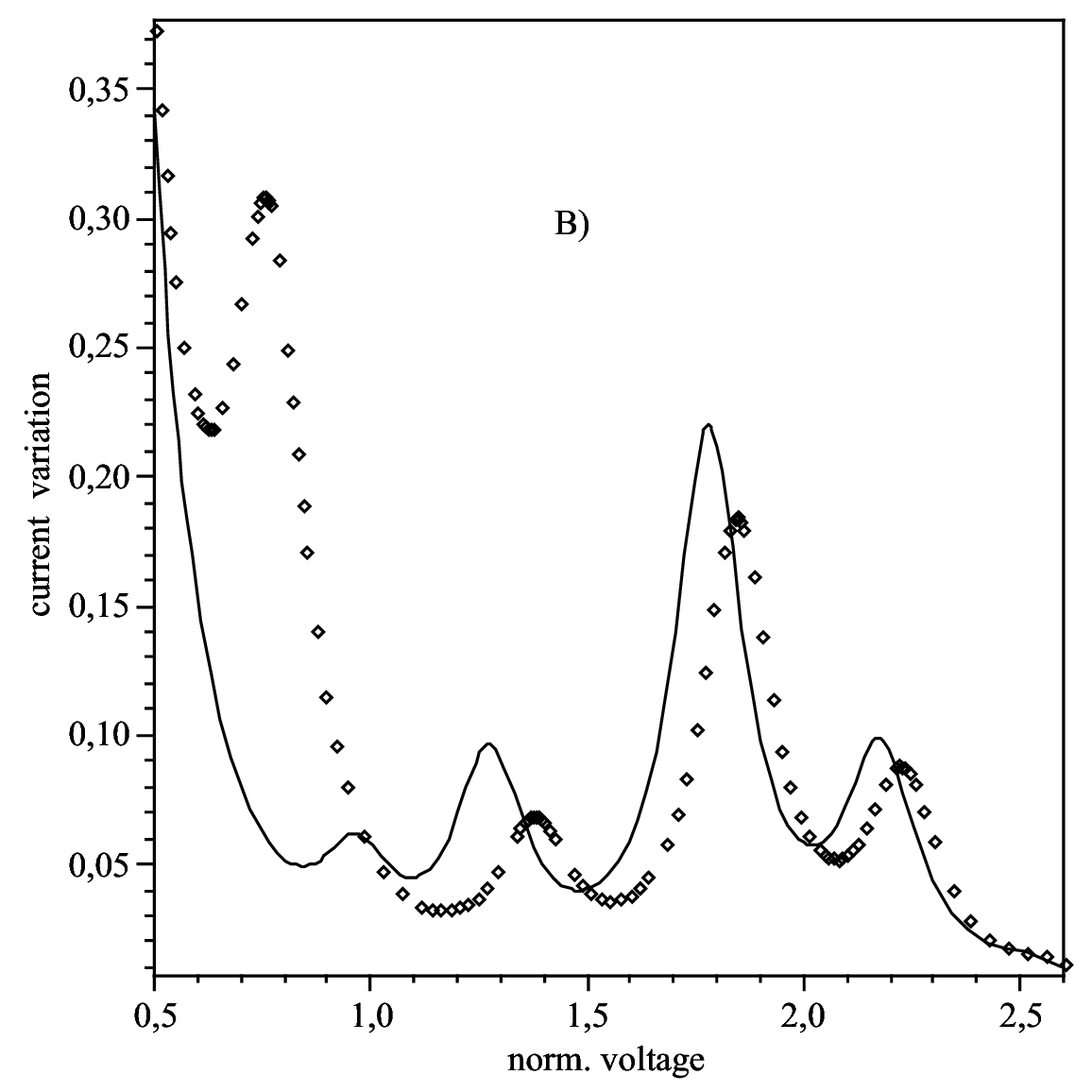}
\end{center}
\caption{Correction to the I-V characteristics due to interaction of
Josephson oscillations and collective modes. The normalized voltage is
defined as $V_{norm}=\Omega _{V}/\Omega _{J}.$ Solid curves correspond to a
small parameter $s\equiv 2(\tilde{d}_{F}/\protect\beta \protect\lambda %
_{L})=0.05$ (negligible effect of the F layer), point curves correspond to $%
s=0.4.$ The plots are presented for the following parameters: A) $\Omega
_{M}/\Omega _{J}=2,$ $\protect\kappa _{H}l_{J}=2;$ B) $\Omega _{M}/\Omega
_{J}=1,$ $\protect\kappa _{H}l_{J}=4.$ The damping coefficients are $\protect%
\gamma _{J}/\Omega _{J}=0.2,\protect\gamma _{M}/\Omega _{J}=0.1.$ }
\end{figure}

The collective modes in the Josephson junctions can be excited by the
internal Josephson oscillations. It is well known that the interaction of
Josephson oscillations with plasma-like waves in a tunnel Josephson junction
in the presence of a weak external magnetic field leads to Fiske steps on
the I-V characteristics (see, e.g., Refs. \cite{Barone}, \cite{EckKulik}).
In the next paragraph we consider the modification of Fiske steps in tunnel
SFIFS or SIFS junctions due to the hybridization of the collective modes
using Eqs. (\ref{Phase}, \ref{Precession}).

b) \textit{Fiske steps}. We assume again that the magnetization $M$\ is
oriented perpendicular to the interfaces and a small external magnetic field
$H_{e}$ is applied parallel to the films. It is important to have in mind
that the magnetization $M_{0}$ is of the order of $100\mathit{Gauss}$ or
larger, whereas $H_{e}$ is about a few $\mathit{Gauss}$. Therefore one can
neglect the in-plane magnetization $M_{y}=H_{e}/4\pi \beta $ in comparison
with $M_{0}$. We assume also that the length of the junction along the $x-$%
axis is shorter than the Josephson length $l_{J}$.

In this limit the phase $\varphi $ can be represented in the form (see Refs. %
\cite{EckKulik}): $\varphi =\varphi _{0}(x,t)+\psi (x,t)$ with $\varphi
_{0}(x,t)=\Omega _{V}t+\kappa _{H}x$ and $\Omega _{V}=2eV/\hbar ,$ $\kappa
_{H}=4\pi \lambda _{L}H_{e}/\Phi _{0}$. Linearizing Eq. (\ref{Phase}) we
obtain the equation for $\psi $

\begin{equation}
\Omega _{J}^{-2}(\frac{\partial ^{2}\psi }{\partial t^{2}}\mathbf{+}\gamma
\frac{\partial \psi }{\partial t})-l_{J}^{2}\frac{\partial ^{2}\psi }{%
\partial x^{2}}=-\sin \varphi _{0}-\frac{c\tilde{d}_{F}}{2\lambda _{L}j_{c}}%
\frac{\partial m_{y}}{\partial x}  \label{Psi}
\end{equation}

The \textit{dc }current is $\eta =(2eV\gamma /\hbar )/\Omega _{J}+\langle
\psi (x,t)\cos \varphi _{0}(x,t)\rangle ,$ where the angular brackets denote
the averaging in space and time. The derivative $\partial m_{y}/\partial x$
can be found from Eq. (\ref{Precession})\ with $\varphi $ replaced by $\psi
. $ The first term on the right-hand side in Eq. (\ref{Psi}) plays a role of
an external force oscillating in space and time and acting on a resonance
system. This equation should be solved using boundary conditions $\partial
\psi /\partial x=0$ at $x=\pm L_{x}/2.$ Carrying out these calculations we
obtain a contribution $\delta \eta \equiv \langle \psi (x,t)\cos \varphi
_{0}(x,t)\rangle $ to the current originating from the collective modes

\begin{equation}
\delta \eta =\frac{1}{2}Im\{\frac{1}{\mathit{D}}[1-\frac{\theta _{H}^{2}}{%
\theta _{V}}\frac{\cos (2\theta _{V})-\cos (2\theta _{H})}{(\theta
_{H}^{2}-\theta _{V}^{2})\sin (2\theta _{V})}]\}  \label{IV}
\end{equation}%
where $\mathit{D}=(\theta _{V}^{2}-\theta _{H}^{2})a_{V}^{2}/\theta
_{V}^{2},a_{V}^{2}=(\Omega _{V}^{2}-i\gamma _{J}\Omega _{V})/\Omega _{J}^{2}$%
, $\theta _{H}=\kappa _{H}L_{x}/2,\theta _{V}=\kappa _{V}L_{x}/2,\kappa
_{V}=a_{V}/l_{V},$ $l_{V}=l_{J}[1+s\Omega _{M}^{2}/(\Omega _{V}^{2}-\Omega
_{Ms}^{2})]^{1/2},\Omega _{Ms}^{2}=\Omega _{M}^{2}(1+s).$ For simplicity we
consider the\textbf{\ }limit $l_{J}>>l_{M}.$

In Fig.2 we plot the dependence $\delta \eta (V)$ vs the normalized voltage $%
\Omega _{V}/\Omega _{J}$ for different values of the parameter $s=2\pi
\tilde{d}_{F}/(\beta \lambda _{L})$. In order to avoid a divergence, we took
into account a finite damping in the F layer replacing $(\Omega
_{V}^{2}-\Omega _{Ms}^{2})$ by $(\Omega _{V}(\Omega _{V}-i\gamma
_{M})-\Omega _{Ms}^{2})$.

In the limit of small values of $s$ ($s\rightarrow 0$) this dependence
describes Fiske steps in the conventional Josephson SIS junction. Solid
lines in Figs. 2 correspond to small values of $s$ ($s=0.05$) when the
influence of the F layer is negligible and point lines correspond to larger
values of $s$ ($s=0.4$). One can see from Fig. 2a that near the frequency of
the magnetic resonance, $\Omega _{V}\approx \Omega _{Ms}=2\Omega _{J}$ there
is a double peak (point curve), which is absent in junctions with a small $s$
(solid curve). Fig. 2b shows that the presence of the F layer leads not only
to a shift of maxima in the dependence $\delta \eta (V)$ but also to a
change of the overall form of this dependence.

To conclude, we have developed theory describing electrodynamics of the
SFIFS or SIFS tunnel Josephson junctions. In the framework of our approach
dynamics of these systems is fully described by Eqs. (\ref{Phase}, \ref%
{Precession}). Solving these equations we found new effects related to
hybridization of charge and spin degrees of freedom. We demonstrated that
the different branches - Josephson plasma-like and spin wave branch - of the
spectrum repel each other. The hybridization of the collective modes results
in new interesting peculiarities of the $I-V$ characteristics. We have
demonstrated that the positions, shape and number of the Fiske steps arising
in the presence of a weak external in-plane magnetic field and an applied
voltage $V$ change. In particular, new peaks appear due to excitation of
magnetic modes.\ We believe that on the basis of the SIFS junctions one may
construct a new type of spectrometers that would allow one to extract an
information about spin excitations in thin ferromagnetic layers\textbf{.} It
seems that an experimental observation of the effects predicted here is not
very difficult and can be performed by measuring $I-V$ characteristics,
spectra of collective excitations, etc., using standard experimental methods
developed for studying Josephson junctions.

We thank SFB 491 for financial support.

\bigskip


\bigskip

\begin{thebibliography}{99}
\bibitem{Josephson} B. D. Josephson, Rev. Mod. Phys., \textbf{36}, 216-220
(1964).

\bibitem{Barone} A.Barone and G.Paterno, \textit{Physics and applications of
the Josephson effect}, Wiley, NY (1982).

\bibitem{Koshelev} A.E. Koshelev and L.N. Bulaevskii, Phys. Rev. B \textbf{77%
}, 014530 (2008).

\bibitem{GolubovRMP} A. A. Golubov, M. Yu. Kupriyanov, and E. Il'ichev, Rev.
Mod. Phys. \textbf{76}, 411 (2004)

\bibitem{BuzdinRMP} A. Buzdin, Rev. Mod. Phys. \textbf{77}, 935 (2005).

\bibitem{BVErmp} F.S. Bergeret, A.F. Volkov, K.B. Efetov, Rev. Mod. Phys.
\textbf{77}, 1321 (2005).

\bibitem{Lyuksyutov} I.F. Lyuksyutov and V.L. Pokrovsky, Adv. Phys. \textbf{%
54}, 67 (2005).

\bibitem{EschrigR} M. Eschrig, T. Lofwander, T. Champel, J. C. Cuevas, J.
Kopu, Gerd Sch\"{o}n, J. Low Temp. Phys. \textbf{147}, 457, (2007).

\bibitem{acJos} X. Waintal, P. W. Brouwer, Phys. Rev. B \textbf{65}, 054407
(2002); I. V. Bobkova and A. M. Bobkov, \textit{ibid} B \textbf{74}, 220504
(2006); J. Michelsen, V. S. Shumeiko, and G. Wendin, Phys. Rev. B \textbf{77}%
, 184506 (2008); Jian-Xin Zhu, Z. Nussinov, A. Shnirman, and A. V. Balatsky,
Phys. Rev. Lett. \textbf{92}, 107001 (2004); E. Zhao and J. A. Sauls, ibid.
\textbf{98}, 206601 (2007);

\bibitem{Braude} V. Braude and Ya. M. Blanter, Phys. Rev. Lett. \textbf{100}%
, 207001 (2008); M. Houzet, \textit{ibid} \textbf{101}, 057009 (2008); F.
Konschelle and A. Buzdin, \textit{ibid }\textbf{102}, 017001 (2009).
.

\bibitem{Maekawa} S. Hikino, M. Mori, S. Takahashi, S. Maekawa, J. Phys.
Soc. Jpn. \textbf{77}, 053707 (2008).

\bibitem{ExpDyn} I.A. Garifullin et al., Appl. Magn. Res. \textbf{22}, 439
(2002); C. Bell, S. Milikisyants, M. Huber, and J. Aarts, Phys. Rev. Lett.
\textbf{100}, 047002 (2008).

\bibitem{Weides} A.S. Vasenko, A.A. Golubov, M.Yu. Kupriyanov, and M.Weides,
Phys. Rev. B \textbf{77}, 134507 (2008); J. Pfeiffer et al., \textit{ibid}
\textbf{77}, 214506 (2008).

\bibitem{Sonin} V. Braude and E.B. Sonin, Phys. Rev. Lett. \textbf{93},
117001 (2004).

\bibitem{AV} A.F. Volkov and A. Anishchanka, Phys. Rev. B \textbf{71},
024501 (2005).

\bibitem{EckKulik} R. E. Eck, D. J. Scalapino, and B. N. Taylor, Phys. Rev.
Lett., \textbf{13,} 15 (1964); I.O. Kulik, JETP Lett., \textbf{2}, 134 (1965).
\end{thebibliography}
\end{document}